\documentclass[12pt]{article}
\usepackage[colorlinks=true,linkcolor=blue,urlcolor=blue,filecolor=black,citecolor=red,pdfstartview=FitV,pdftitle={},pdfsubject={},
pdfkeywords={},pdfpagemode=None,bookmarksopen=true]{hyperref}
\usepackage{graphicx}%Include figure filespp
\usepackage{epstopdf}%
\usepackage{amsmath}
\usepackage{amsfonts}
\usepackage{amssymb}
\usepackage{color}%
\usepackage{dcolumn}% Align table columns on decimal pointppp
\usepackage{slashed}
\usepackage{amssymb,ulem}
\usepackage{float}
\usepackage{amsthm,amsmath,amssymb}
\usepackage{mathrsfs}
\usepackage{subfigure}
\usepackage{wrapfig}
\usepackage{indentfirst}
\providecommand{\U}[1]{\protect\rule{.1in}{.1in}}

\begin{document}

\vspace{12mm}

\begin{center}
{{{\Large {\bf Spin-induced scalarization of Kerr-Newman black holes in Einstein-Maxwell-scalar theory}}}}\\[10mm]

{Meng-Yun Lai$^a$\footnote{mengyunlai@jxnu.edu.cn;},
Yun Soo Myung$^b$\footnote{ysmyung@inje.ac.kr;},
Rui-Hong Yue$^c$\footnote{Corresponding author. rhyue@yzu.edu.cn;} \\
and De-Cheng Zou$^{c}$\footnote{Corresponding author. dczou@yzu.edu.cn;}}\\[8mm]

{${}^a$College of Physics and Communication Electronics, Jiangxi Normal University, Nanchang 330022, China\\[0pt] }
{${}^b$Institute of Basic Sciences and Department  of Computer Simulation,\\ Inje University Gimhae 50834, Korea\\[0pt] }
{${}^c$Center for Gravitation and Cosmology and College of Physical Science and Technology, Yangzhou University, Yangzhou 225009, China\\[0pt]}
\end{center}

\vspace{2mm}
\vspace{2mm}

\begin{abstract}

In this paper, we consider the tachyonic instability of Kerr-Newman (KN) black holes in the Einstein-Maxwell-scalar (EMS) theory with nonminimal negative scalar coupling to Maxwell term, and then obtain a bound  $(\frac{a}{r_+}\geq0.4142)$ for spin parameter $a$ of KN black hole in the limit of coupling constant $\alpha\rightarrow-\infty$ by using analytical
method. In addition, we perform a $(2+1)$-dimensional time evolution of linearized scalar field perturbation on the KN black hole background by implementing the hyperboloidal foliation method numerically in the EMS theory. It recovers the \textit{spontaneous scalarization} phenomenon for KN black hole in some certain regions of the parameter spaces. We further plot corresponding threshold curves $\alpha(a)$ which describe a boundary between bald KN black holes and scalarized spinning black holes in the EMS theory.
\end{abstract}
\vspace{5mm}

\newpage
\renewcommand{\thefootnote}{\arabic{footnote}}
\setcounter{footnote}{0}
\section{Introduction}

The no-hair theorem states that a black hole can be completely characterized by only three externally observable classical parameters: mass, electric charge, and angular momentum in general relativity 
\cite{Carter:1971zc,Ruffini:1971bza}.
J. D. Bekenstein et al. also found that the asymptotically flat spacetime cannot admit any hairy
black hole solutions \cite{Bekenstein:1974sf,Bekenstein:1975ts}, since the scalar field was divergent on the horizon and stability analysis showed that they were unstable \cite{Bronnikov:1978mx} in general relativity . However, this theorem suffers from challenges due to the existence of hairy black holes possessing extra macroscopic degrees of freedom in other theories. For instance, \textit{spontaneous scalarization} is considered as a new mechanism to obtain a black hole with scalar hair, since the original proposal \cite{Damour:1993hw}.
It occurs in the context of scalar-tensor theories possessing the scalar field nonminimally coupled to either Gauss-Bonnet term~\cite{Doneva:2017bvd,Silva:2017uqg,Antoniou:2017acq}, or to Maxwell term~\cite{Herdeiro:2018wub}-\cite{Myung:2018vug}, where the scalar field results in destabilization of scalar-free black holes and to form scalarized (charged) black holes. Then, the phenomenon of tachyonic instability happens in the static spherical case.

Recently, some authors further investigated the curvature-induced spontaneous scalarization
of Kerr black holes branching out of the Schwarzschild solution in the Einstein-scalar-Gauss-Bonnet (ESGB)
theory with a positive coupling parameter~\cite{Cunha:2019dwb,Collodel:2019kkx}.
Later on, an $a$-bound of $a/M \ge 0.5$ was found as the onset of scalarization for
Kerr black holes using $(1+1)$-dimensional simulations~\cite{Dima:2020yac}.
This critical rotation parameter $\left(a/M\right)_{crit}\simeq0.5$ was also computed analytically~\cite{Hod:2020jjy} and
numerically~\cite{Zhang:2020pko,Doneva:2020nbb}, and it was found that the low rotation suppresses spontaneous scalarization.
Doneva \textit{et al.}~\cite{Doneva:2020kfv} also considered the impact of massive scalar field perturbation for Kerr black hole
in the ESGB gravity. Furthermore, the spin-induced scalarized black holes were numerically
constructed for high rotation and negative coupling parameter in the ESGB gravity~\cite{Herdeiro:2020wei,Berti:2020kgk}.

In this paper, we first turn to explore, using analytical techniques, the \textit{spontaneous scalarization} of Kerr-Newman (KN) black holes in the Einstein-Maxwell-scalar (EMS) theory. Considering the nonseparability problem and ``outer boundary problem" \cite{Zenginoglu:2011zz,Thuestad:2017ngu} in the numerical calculations with the traditional method, we will adopt $(2+1)$-dimensional hyperboloidal foliation method~\cite{Zenginoglu:2007jw}-\cite{Zenginoglu:2010cq} to deal with the time evolution equation of scalar field perturbation on the KN black hole background, and expect to get a complete and objective picture about the stability and spontaneous scalarization of KN black hole in the EMS gravity.

This paper is organized as follows. In Sec.~\ref{2s}, we introduce the EMS theory and derive relevant perturbed field equations. In Sec.~\ref{5s},
we study the onset of spontaneous scalarization phenomenon for KN black holes by using the analytical method. The numerical method employed is described in Sec.~\ref{3s}, and the main results are presented in Sec.~\ref{4s}. Section.~\ref{6s} is devoted to summary and discussions.

\section{EMS gravity and perturbed field equations}\label{2s}

We start with the EMS gravity theory as
\begin{eqnarray}
\mathcal{S}_{\rm EMS}=\frac{1}{16\pi}\int d^4x{\sqrt{-g}\Big(R-2\partial_\mu\phi\partial^\mu\phi-f(\phi)F_{\mu\nu}F^{\mu\nu}\Big)}, \label{action}
\end{eqnarray}
where $R$ is the Ricci scalar, $F_{\mu\nu}=\partial_{\mu}A_{\nu}-\partial_{\nu}A_{\mu}$ is the Maxwell field and $\phi$ is the scalar
field. The coupling function $f(\phi)$ controls the nonminimal coupling of scalar field $\phi$ to the Maxwell term $F^2\equiv F_{\mu\nu}F^{\mu\nu}$.

From the action \eqref{action}, the three field equations can be easily obtained with respect to the
field variables $g_{\mu\nu}$, $\phi$, and $A_\mu$
\begin{eqnarray}
&&R_{\mu\nu}-\frac{1}{2}R g_{\mu\nu}=2\partial _\mu \phi\partial _\nu \phi-(\partial \phi)^2g_{\mu\nu}
+2f(\phi)\Big(F_{\mu\rho}F_{\nu}~^\rho-\frac{F^2}{4}g_{\mu\nu}\Big), \label{g-eql}\\
&&\square \phi -\frac{1}{4}\frac{d f(\phi)}{d\phi}F^2=0, \label{s-equa1}\\
&&\partial_\mu\left(\sqrt{-g}f(\phi)F^{\mu\nu}\right)=0\label{M-eq1}.
\end{eqnarray}

When selecting the no-scalar hair (by denoting the overbar)
\begin{eqnarray}
\bar{\phi}=0,
\end{eqnarray}
the action (\ref{action}) and corresponding field equations \eqref{g-eql}-\eqref{M-eq1}  should reduce to the Maxwell's theory and the coupling function $f(\bar{\phi})$ is obtained as $f(0)=1$. This theory possesses the axisymmetric KN spacetime for rotation metric ansatz. The metric in the Boyer-Lindquist coordinates reads as
\begin{eqnarray}
ds^2_{\rm KN} &\equiv& \bar{g}_{\mu\nu}dx^\mu dx^\nu=-\frac{\Delta-a^2\sin^2\theta}{\rho^2}dt^2
-\frac{2a\sin^2\theta(r^2+a^2-\Delta)}{\rho^2}dt d\varphi\nonumber\\
&&+\frac{[(r^2+a^2)^2-\Delta a^2 \sin^2\theta]\sin^2\theta}{\rho^2} d\varphi^2+ \frac{\rho^2}{\Delta}dr^2 +\rho^2 d\theta^2, \label{KN-sol}
\end{eqnarray}
with
\begin{eqnarray}
 \Delta= r^2-2Mr+a^2+Q^2,\quad \rho^2=r^2+a^2\cos^2\theta,\quad a=\frac{J}{M}.\nonumber
\end{eqnarray}
 The corresponding vector potential is
\begin{eqnarray}
 \bar{A}=-\frac{Qr}{\rho^2}\left(dt-a\sin^2\theta d\varphi\right).
\end{eqnarray}
The outer and inner horizons are found  by demanding $\Delta=(r-r_+)(r-r_-)=0$ as
\begin{eqnarray}
r_{\pm}=M\pm \sqrt{M^2-a^2-Q^2}.\label{radius}
\end{eqnarray}
In this background, the Maxwell term is obtained as
\begin{eqnarray}
  \bar{F}^2=-\frac{2Q^2(r^4-6a^2r^2\cos^2\theta+a^4\cos^4\theta)}{\left(r^2+a^2\cos^2\theta\right)^4}. \label{b-maxwell}
\end{eqnarray}

We now consider generic linear perturbations of the
KN black hole solution
\begin{equation}
g_{\mu\nu}=\bar{g}_{\mu\nu}+h_{\mu\nu},\quad\phi=\bar{\phi}(=0)+\delta \phi, \quad F_{\mu\nu}=\bar{F}_{\mu\nu}+{\cal F }_{\mu\nu} \label{lin-eqs}
\end{equation}
with the linearized field tensor
\begin{equation}
{\cal F }_{\mu\nu} =\partial_\mu a_\nu-\partial_\nu a_\mu.
\end{equation}
Substituting Eq.~(\ref{lin-eqs}) into field equations \eqref{g-eql}-\eqref{M-eq1}, we can obtain the perturbed equations
 \begin{eqnarray}
&&\delta G_{\mu\nu}(h)=2 \delta T^{\rm M}_{\mu\nu}, \label{per-eq1}\\
&&\bar{\nabla}^\mu {\cal F }_{\mu\nu}=0,\label{per-eq2}\\
&&\Big(\bar{\square}-\mu^2_{\rm eff}\Big)\delta \phi=0,\label{per-eq3}
\end{eqnarray}
where
\begin{eqnarray}
  \delta G_{\mu\nu} &=& \delta R_{\mu\nu}-\frac{1}{2} \bar{g}_{\mu\nu} \delta R -\frac{1}{2} \bar{R} h_{\mu\nu}, \label{l-G} \\
  \delta T^{\rm M}_{\mu\nu} &=&\bar{F}_{\nu}~^\rho {\cal F }_{\mu\rho}+\bar{F}_{\mu}~^\rho {\cal F }_{\nu\rho}-\bar{F}_{\mu\rho}\bar{F}_{\nu\sigma}h^{\rho\sigma} \label{lin-max} \\
                    &+&\frac{1}{2}(\bar{F}_{\kappa\eta}{\cal F }^{\kappa\eta}-\bar{F}_{\kappa \eta}\bar{F}^\kappa~_\sigma h^{\eta\sigma})\bar{g}_{\mu\nu}-\frac{1}{4}\bar{F}^2h_{\mu\nu},\nonumber \\
  \mu_{\rm eff}^2&=&\frac{\bar{F}^2}{4}\frac{d^2 f(\phi)}{d\phi^2}(0) \label{lin-mass}.
  \end{eqnarray}
In analyzing the instability of the KN black hole in the EMS theory, we first consider the  linearized equations \eqref{per-eq1} and \eqref{per-eq2}
because two perturbations of metric $h_{\mu\nu}$ and  vector $a_{\mu}$ are coupled to each other.
These are the same as  those for the Einstein-Maxwell theory~\cite{chandra}.
Recently, it was shown  that the KN black hole is  stable against metric
and vector perturbations in the Einstein-Maxwell theory~\cite{Dias:2015wqa}.
This implies that one does  not need to solve Eqs.~\eqref{per-eq1} and \eqref{per-eq2} for the instability analysis in the EMS theory. Therefore, we focus on solving the the linearized scalar equation \eqref{per-eq3} numerically which may determine
the tachyonic instability of KN black hole in the EMS theory.

For the form of coupling function $f(\phi)$, we can adopt a similar quadratic coupling function $1+\alpha\phi^2$, or
exponential coupling function $e^{\alpha\phi^2}$, or etc., which satisfy the condition $f(0)=1$ above. Moreover, these coupling functions were also used in discussions for the tachyonic instability of charged scalarized static black holes in the EMS theory~\cite{Herdeiro:2018wub,Fernandes:2019rez,Myung:2018vug}. Then, we have
\begin{eqnarray}\label{fphi}
f(0)=1,\quad \frac{d f}{d\phi}(0)=0, \quad \frac{d^2f}{d\phi^2}(0)=2\alpha.
\end{eqnarray}
From Eqs.~\eqref{b-maxwell}, \eqref{lin-mass}, and \eqref{fphi}, the effective mass squared of perturbed scalar field can be obtained as
\begin{eqnarray}
\mu_{\rm eff}^2&=&\frac{\bar{F}^2}{4}\frac{d^2 f(\phi)}{d\phi^2}(0)=\frac{\alpha \bar{F}^2 }{2}\nonumber\\
&=&-\frac{\alpha Q^2(r^4-6a^2r^2\cos^2\theta+a^4\cos^4\theta)}{\left(r^2+a^2\cos^2\theta\right)^4}.\label{effmass}
\end{eqnarray}

For charged static case [Reissner-Nordstr\"{o}m (RN) black hole] background, the effective mass of perturbed scalar field is given by
\begin{eqnarray}\label{staticmass}
\mu_{\rm eff}^2=-\frac{\alpha Q^2}{r^4}.
\end{eqnarray}
Notice that Refs.~\cite{Herdeiro:2018wub,Fernandes:2019rez,Myung:2018vug} have recovered that tachyonic instability of RN black hole is  only characterized by the presence of an
effective negative (squared) mass term $\mu_{\rm eff}^2<0$ with positive coupling parameter $\alpha>0$, but the positive effective mass term $\mu_{\rm eff}^2>0(\alpha<0)$ suppresses the spontaneous scalarization of RN black hole in the EMS gravity.

In the next sections, we turn to investigate the onset of spontaneous scalarization in the KN black hole spacetimes of
EMS theory. We expect to observe the object pictures of the spontaneous scalarization phenomenon under the influences of coupling parameter $\alpha$, rotation parameter $a$, mass $M$, and charge $Q$ of KN black hole in the EMS gravity.

\section{Analytical results}\label{5s}

Recently, Hod~\cite{Hod:2020jjy} has obtained the critical rotation parameter
\begin{eqnarray}
a_{crit}\equiv\left(\frac{a}{M}\right)_{crit}\simeq0.505,
\end{eqnarray}
which marks the boundary between bald Kerr black holes and hairy (scalarized) spinning black holes in the ESGB gravity with negative values of parameter $\alpha<0$. Later, we also introduced this analytic approach to investigate the critical rotation parameter $(a_{crit}\equiv\left(a/M\right)_{crit})$ of tachyonic instability condition of Kerr black holes in the Einstein-scalar-Chern-Simons gravity with negative coupling parameter~\cite{Myung:2020etf}. In this section, we try to derive the critical rotation parameter $(a_{crit})$ for KN black hole in the EMS gravity by using the analytical techniques.

As shown in Ref. \cite{Dima:2020rzg}, we can also introduce coordinates transformations
\begin{eqnarray}
d\varphi^*=d\varphi +\frac{a}{\Delta} dr,\quad dx=\frac{r^2+a^2}{\Delta} dr,
\end{eqnarray}
we can obtain the following explicit expression
for the Klein-Gordon equation \eqref{per-eq3} in the coordinates
\begin{eqnarray}
% \nonumber % Remove numbering (before each equation)
  &&\left[ (r^2+a^2)^2-
      \Delta a^2\sin^2\theta\right]\partial^2_t \delta\phi
    -(r^2+a^2)^2\partial_x^2\delta\phi
    -2r\Delta\partial_x\delta\phi\nonumber\\
    &&+2a\left(2Mr-Q^2\right)\partial_t\partial_{\varphi^*}\delta\phi
    -2a(r^2+a^2)\partial_x\partial_{\varphi^*}\delta\phi
   \nonumber\\
    &&-\Delta\left[\frac{1}{\sin\theta}\partial_\theta(\sin\theta\partial_\theta\delta\phi)
    +\frac{1}{\sin^2\theta}\partial^2_{\varphi^*}\delta\phi\right]\nonumber\\
    &&+\Delta\left(r^2+a^2\cos^2\theta\right)\mu_{eff}^2\delta\phi=0. \label{perturbedEq2-1}
\end{eqnarray}

Let us introduce a projection of linearized scalar equation \eqref{perturbedEq2-1} onto a basis of spherical harmonics. Based on a decomposition of the scalar field in a series of spherical harmonics $Y_{lm}(\theta,\varphi)$
\begin{eqnarray}
 \delta \phi(t,r,\theta,\varphi) =\sum_{m}\sum_{l=|m|}^{\infty} \frac{\psi_{l m}(t,r)}{r} Y_{l m}(\theta,\varphi),
\end{eqnarray}
we can insert this decomposition into Eq.~\eqref{perturbedEq2-1} and obtain the coupled $(1+1)$-dimensional evolution equation
\begin{eqnarray}
&&\Big[(r^2+a^2)^2-a^2\Delta(1-c^m_{ll})\Big]\ddot{\psi}_{l m}+
a^2\Delta\left(c^m_{l,l+2}\ddot{\psi}_{l+2,m}+c^m_{l,l-2}\ddot{\psi}_{l-2,m}\right)\nonumber\\
&&+2i a m \left(2M r-Q^2\right)\dot{\psi}_{l m}-(r^2+a^2)^2\psi''_{l m}-
\Big[2i a m(r^2+a^2)-2a^2\Delta/r\Big]\psi'_{l m}\nonumber\\
&&\Delta\Big[l(l+1)+\frac{2M}{r}-\frac{2a^2+2Q^2}{r^2}+\frac{2i a m}{r}\Big]\psi_{l m}\nonumber\\
&&+\Delta\sum_{j m'}<l m|\mu^2_{\rm eff}(r^2+a^2\cos^2\theta)|j m'>\psi_{j m'}=0,
\end{eqnarray}
Here the prime denotes the derivative of coordinate $x$ and function $\psi_{l m}$ is given by
\begin{eqnarray}
\psi_{l m}\equiv\int r\delta\phi Y^*_{l m}d\Omega.
\end{eqnarray}
Because of axisymmetry, different $m$ modes decouple from one another, but the decomposition in spherical harmonics generates couplings for each $l$ mode to the $l\pm2$ modes.

At the threshold  of tachyonic instability, the mixed term of
$\Delta\sum_{j}< l m | \mu^2_{\rm eff}(r^2+a^2\cos^2\theta)| j m>\psi_{j m}$ can be replaced by a single term as
\begin{equation}
\Delta< l_1 m=0 | \mu^2_{\rm eff}(r^2+a^2\cos^2\theta)| l_2 m=0>\psi_{l_2 m}
\end{equation}
at asymptotically late times.
The onset of spontaneous scalarization is related to  an effective  binding potential well in the near horizon whose two turning points of $r_{\rm in}$ and $r_{\rm out}$ are classified by the relation of $r_{\rm out}\ge r_{\rm in}=r_+$.  A critical black hole with $a=a_{crit}$  denotes the boundary between KN black hole and rotating charged scalarized black hole existing  in the limit of $\alpha \to -\infty$. It is characterized by  the presence of a degenerate  binding potential well whose two turning points
merge at the outer horizon ($r_{\rm out}= r_{\rm in}=r_+$) as
\begin{eqnarray}
< l_1 m=0 | \mu^2_{\rm eff}(r^2+a^2\cos^2\theta)| l_2 m=0>|_{r=r_+(a_{crit})}=0, \quad for \quad a= a_{crit},
\end{eqnarray}
in the limit of $\alpha \to -\infty$.
In this case, the critical rotation parameter $a_{crit}$ is determined by the resonance condition
\begin{eqnarray}
\int_0^\pi\frac{(r_+^4-6a^2r_+^2\cos^2\theta+a^4\cos^4\theta)}{\left(r_+^2+a^2\cos^2\theta\right)^3}Y_{l_10}Y_{l_20} \sin\theta d\theta =0, \quad for \quad a= a_{crit}.\label{res-con}
\end{eqnarray}

To solve Eq.~\eqref{res-con} for $a_{crit}$ analytically, one introduces two new variables
\begin{eqnarray}
\hat{a}\equiv\frac{a_{crit}}{r_+}, \qquad
x=\hat{a} \cos \theta.
\end{eqnarray}
Then, Eq.~\eqref{res-con} takes the form
\begin{eqnarray} \label{con-im}
\int^{\hat{a}}_{-\hat{a}} \frac{(1-6x^2+x^4)}{(1+x^2)^3}  Y_{l_10}(x/\hat{a})Y_{l_20}(x/\hat{a})dx=0.
\end{eqnarray}
In the asymptotic limit of $l_1=l_2 =l\to \infty$, one finds $Y^2_{l0}\to \delta(\theta)$ around the poles of  $\theta=0,\pi(x=\hat{a},-\hat{a})$.
In this case, Eq.~\eqref{con-im} leads to a condition for the  resonance
\begin{eqnarray}
1-6\hat{a}^2+\hat{a}^4 =0.\label{hata}
\end{eqnarray}
Here we consider the smallest possible value of the black hole spin parameter $\hat{a}$ which allows the existence of the nonminimally coupled scalar clouds
\begin{eqnarray}\label{critac}
\hat{a}=0.4142
\end{eqnarray}
for the critical black hole rotation parameter.
However, the other solution $\hat{a}=2.4142$ of Eq.~(\ref{hata}) provides no physical solution for $a_{crit}$.

\begin{figure}[ht]
\centering
\includegraphics[width=0.39\textwidth]{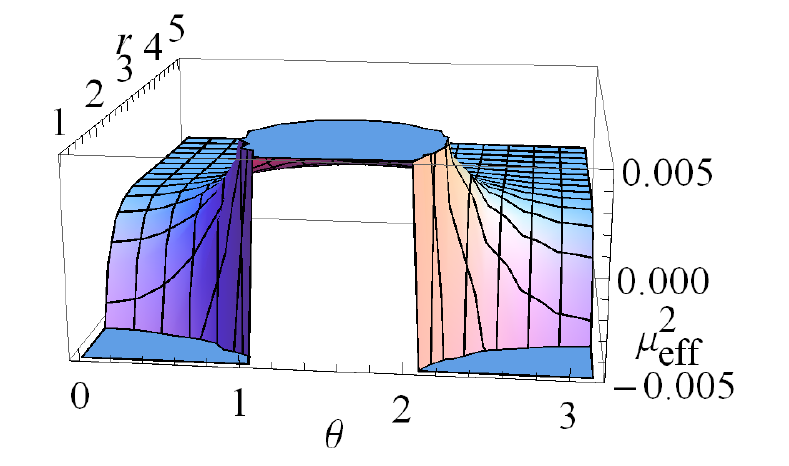}
\quad
\includegraphics[width=0.39\textwidth]{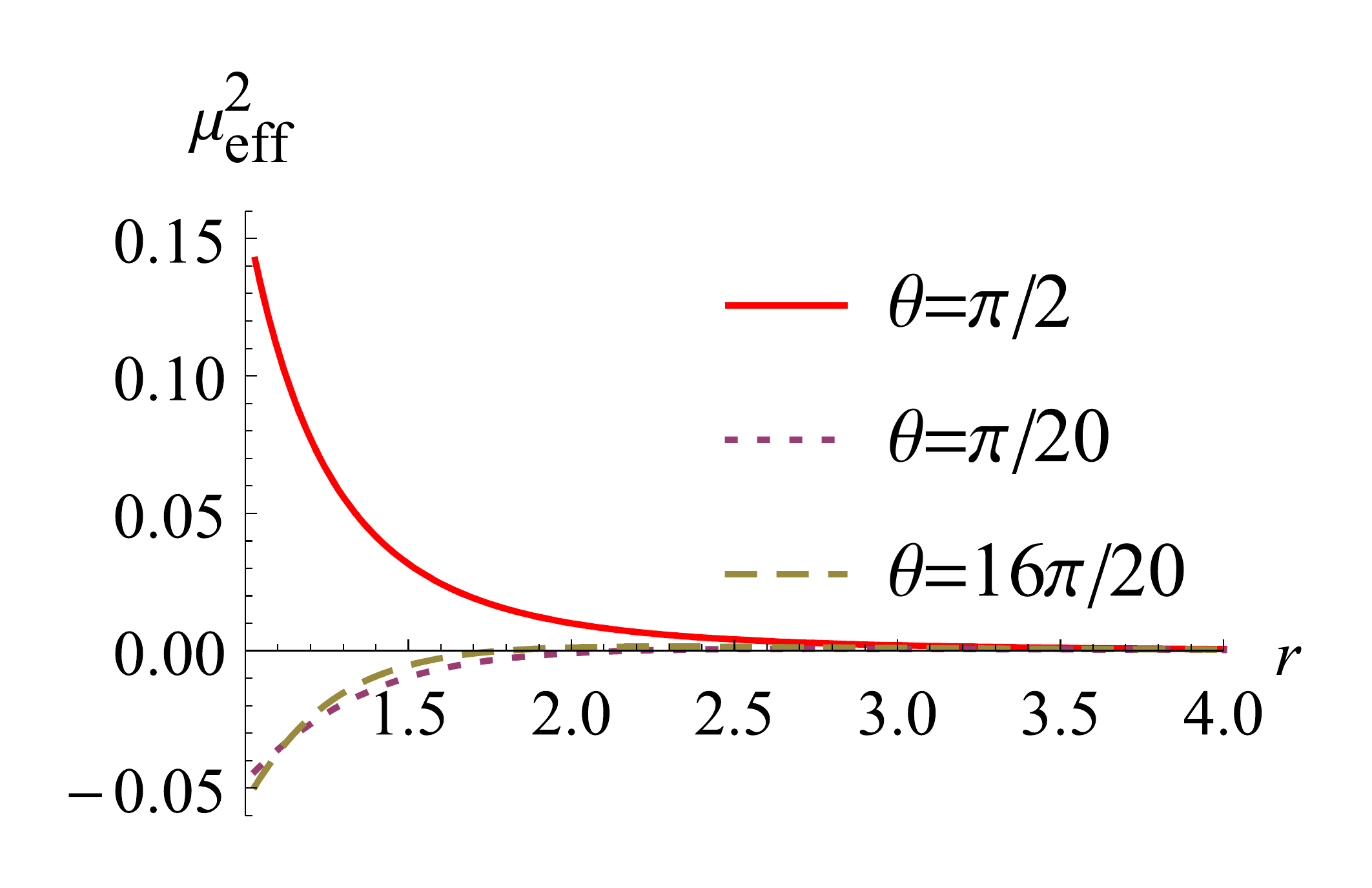}
\caption{ The $3D$ (left) and $2D$ (right) graphs for the effective mass term ($\mu^2_{\rm eff}$) with $M=1$, $Q=0.4$ and $a=0.916$.  }\label{Fig01}
\end{figure}

\begin{figure}[ht]
\centering
\includegraphics[width=0.38\textwidth]{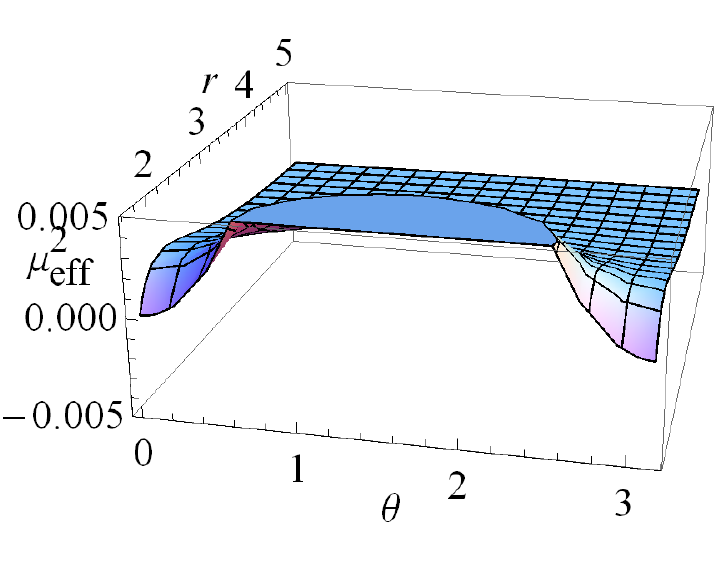}
\quad
\includegraphics[width=0.38\textwidth]{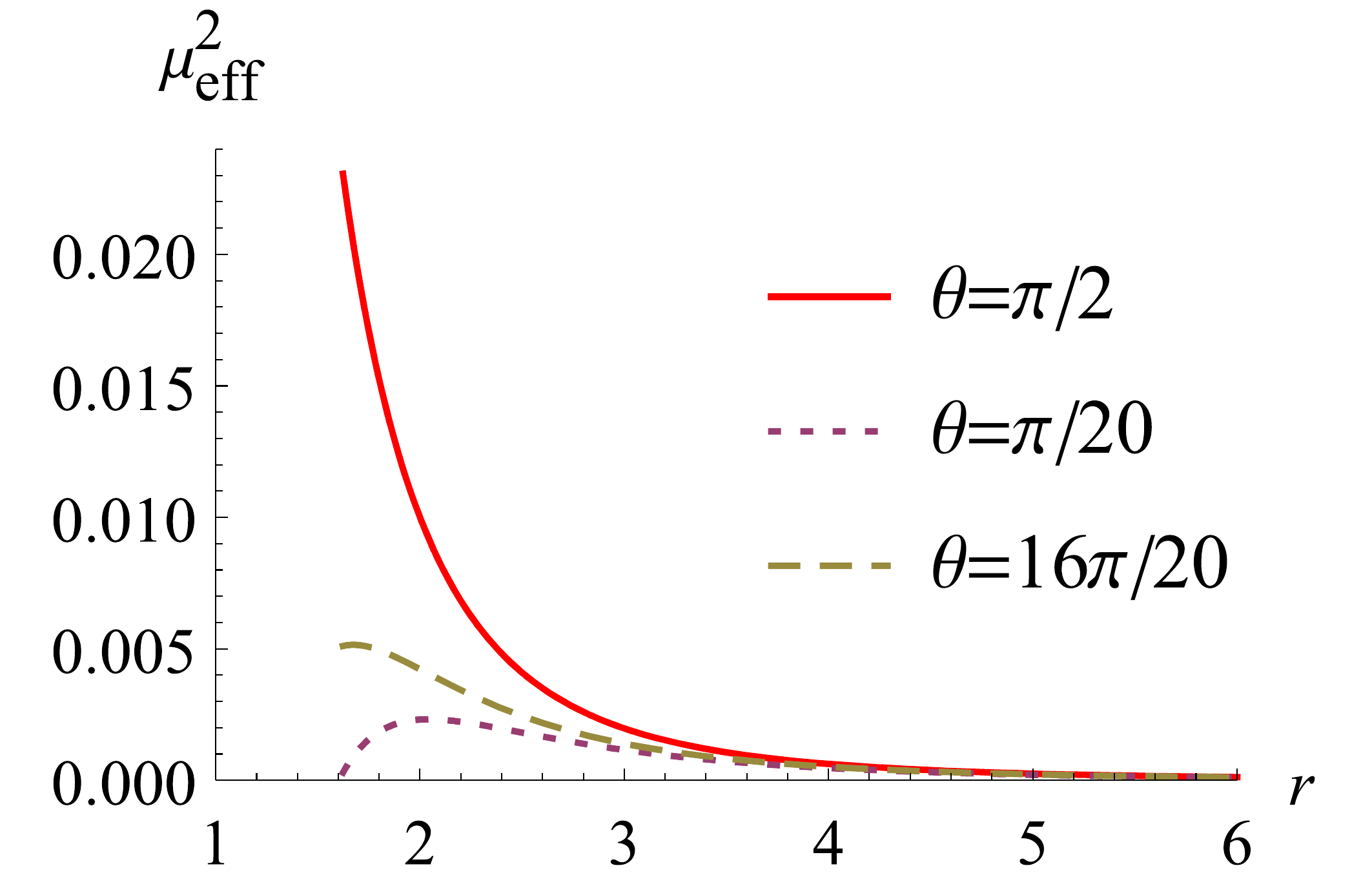}
\caption{ The $3D$ (left) and $2D$ (right) graphs for the effective mass term ($\mu^2_{\rm eff}$) with $M=1$, $Q=0.4$ and $a=0.67$. }\label{Fig02}
\end{figure}

\begin{figure}[ht]
\centering
\includegraphics[width=0.38\textwidth]{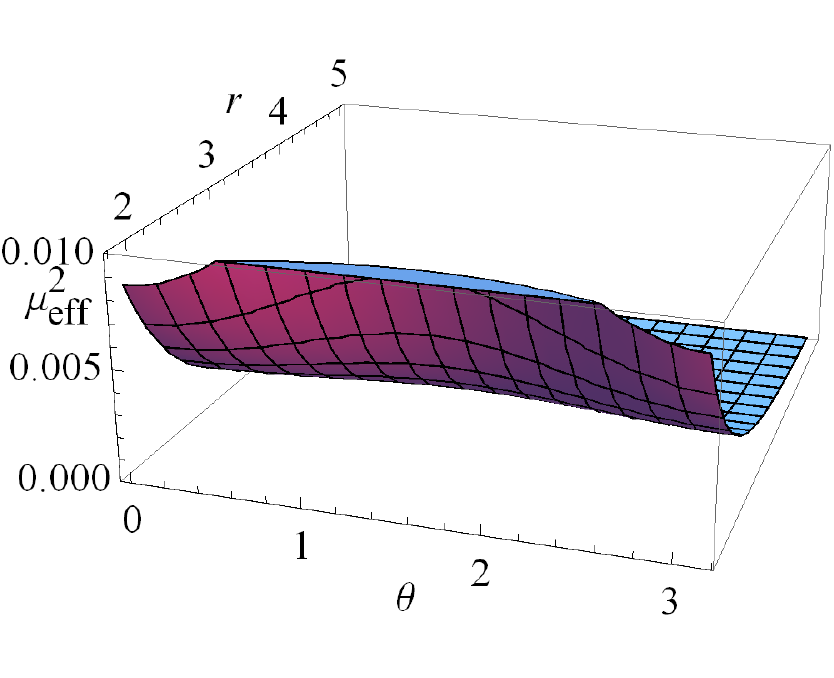}
\quad
\includegraphics[width=0.38\textwidth]{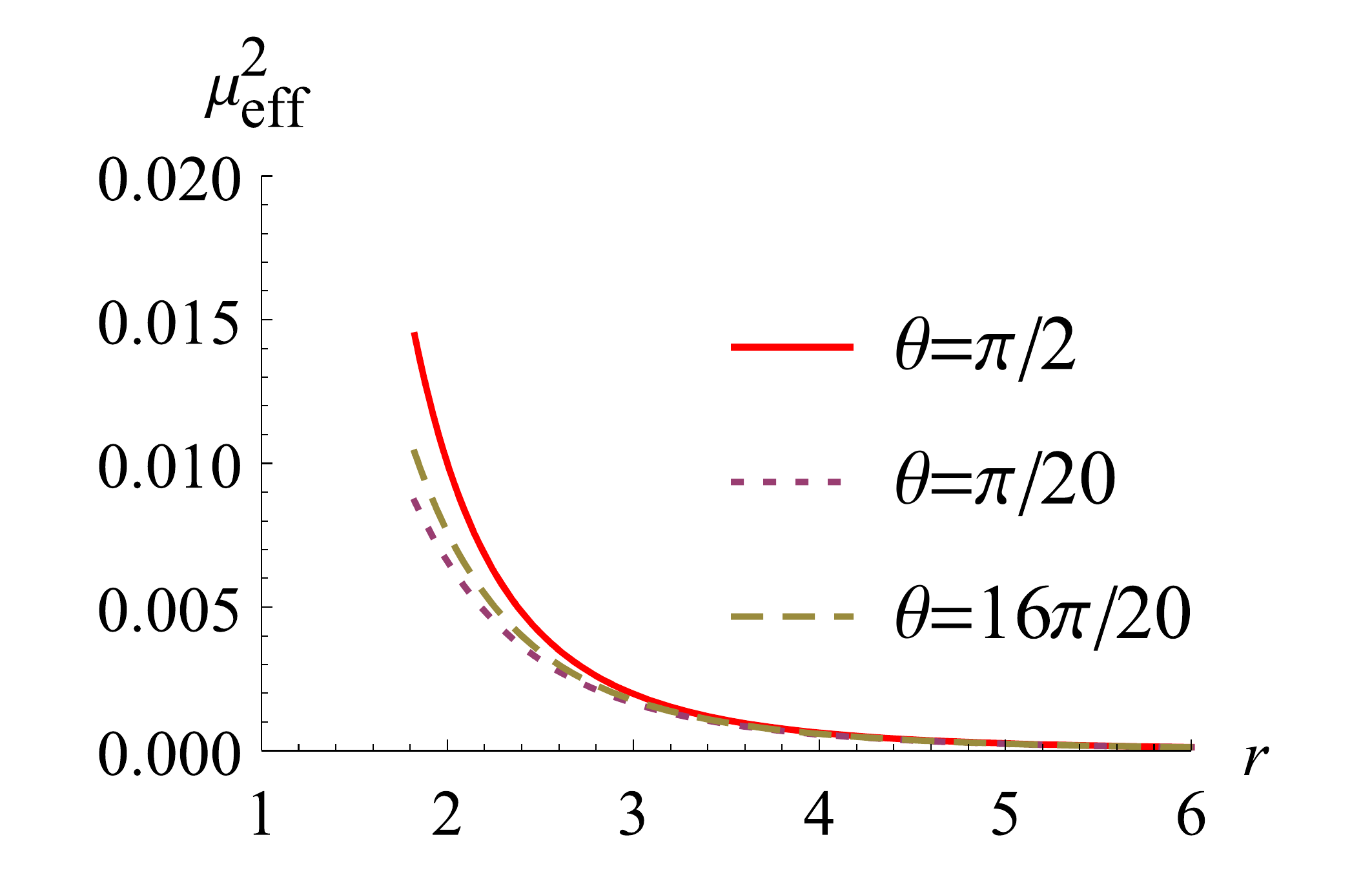}
\caption{ The $3D$ (left) and $2D$ (right) graphs for the effective mass term ($\mu^2_{\rm eff}$) with $M=1$, $Q=0.4$ and $a=0.4$.}\label{Fig03}
\end{figure}

Taking $M=1$ and $Q=0.4$ for example, one case of $\hat{a}=0.4142$ solves to find
the critical rotation parameter \begin{eqnarray}
a_{crit}=\left(\frac{a}{M}\right)_{crit}=0.6722\simeq 0.67
\end{eqnarray}
with the KN black hole radius $r_+$ in Eq.~\eqref{radius}. It implies the $a$ bound
\begin{eqnarray}
a\ge a_{crit} =0.67 \label{abound}
\end{eqnarray}
in the KN black hole background,  being independent of the coupling $\alpha$.
That is, the KN black holes with $a <0.67$ could not develop the tachyonic instability and could not have rotating charged scalarized black holes. Furthermore, one observes from Fig.~\ref{Fig01} with $M=1$, $Q=0.4$, and $\alpha=-1$ that a graph for $a=0.916>a_{crit}=0.67$ represents a positive region around $\theta=\pi/2$ and the negative regions around $\theta=0,\pi$, while the graphs for $a=0.67\simeq a_{crit}$ and $a=0.4<a_{crit}$ show the whole positive regions, see Figs.~\ref{Fig02} and \ref{Fig03}.  We note  that the negative region in the $\theta$ direction inducing spontaneous scalarization decreases to zero as $a$ decreases from $a=0.916$ to $a=a_{crit}$, while positive region around $\theta=\pi/2$ suppressing spontaneous scalarization  increases as $a$ decreases. This confirms that $a_{crit}=0.67$ is considered as the critical case for the tachyonic instability where the coupling parameter $\alpha$ goes to $-\infty$. Also, this implies the $a$ bound \eqref{abound}.

At this stage,  we note that the known $a$-bounds include  $a/M\ge 0.5$ for GB coupling term~\cite{Dima:2020yac},
$a/M\ge 0.26$ for GB+Chern-Simons coupling term~\cite{Zou:2021ybk}, and no $a$-bound for
Chern-Simons coupling term~\cite{Myung:2020etf}
in the Kerr black hole background.
It is worth noting that \eqref{abound} is regarded as the largest bound up to now.
It is clear that the threshold  curve $\alpha=\alpha_{\rm th}(a)$ for KN black holes depends on  $a$ in the EMS gravity.
We conjecture from~\cite{Zou:2021ybk} that $\alpha=\alpha_{\rm th}(a)$ is a rapidly decreasing function of $a$. Its form will be determined by performing  numerical computations in following sections.

\section{Numerical method}\label{3s}

In general, it is not easy to solve this partial differential equation \eqref{per-eq3}. With traditional numerical methods, one has to truncate the infinite radial computational domain to a finite range and put boundary conditions at the outer edges. It inevitably results in spurious wave reflections from the outer edges, namely so-called ``outer boundary problem". In order to overcome this problem, we adopt the (2+1)-dimensional hyperboloidal foliation method to solve Eq.~(\ref{per-eq3}) numerically, where the ingoing (outgoing) boundary condition at the horizon (infinity) is satisfied automatically, so we do not need to worry about the sophisticated outer boundary problem  \cite{Zenginoglu:2011zz,Thuestad:2017ngu}.
In fact, the hyperboloidal foliation method has been also used to discuss the behavior of time evolution for scalar field perturbation on the Kerr black hole in the ESGB~\cite{Zhang:2020pko} and  Einstein-scalar-Chern-Simons~\cite{Gao:2018acg} gravity.
Now we try to calculate the time evolution equation of scalar field perturbation on the KN black hole background in the EMS gravity by using the hyperboloidal foliation method.

This method mainly contains two successive coordinates transformations, which have the advantages that the time slices are horizon penetrating and connect to future null infinity such that no boundary conditions are needed.
First, we introduce the ingoing Kerr-Schild coordinates $\{\tilde{t},r,\theta,\tilde{\varphi}\}$
by considering the coordinate transformations
\begin{eqnarray}
d\tilde{t}=dt+\frac{2 Mr-Q^2}{\Delta} dr,\quad d\tilde{\varphi}=d\varphi +\frac{a}{\Delta} dr.
\end{eqnarray}
and the KN black hole metric \eqref{KN-sol} becomes
\begin{eqnarray}
 ds^2&=&-\left(1-\frac{2M r-Q^2}{\rho^2}\right)d\tilde{t}^2
 -\frac{2a\left(2M r-Q^2\right)\sin^2\theta}{\rho^2}d\tilde{t} d\tilde{\varphi}+\frac{4M r-2Q^2}{\rho^2}d\tilde{t}dr\nonumber\\
 &&+\left(1+\frac{2M r-Q^2}{\rho^2}\right)d r^2-2a\sin^2\theta\left(1+\frac{2M r-Q^2}{\rho^2}\right) dr d\tilde{\varphi} +\rho^2 d\theta^2\nonumber\\
 &&+\left(r^2+a^2+\frac{(2M r-Q^2)a^2\sin^2\theta}{\rho^2}\right)\sin^2\theta d\tilde{\varphi}^2. \label{KN-KS}
\end{eqnarray}

Given the axial symmetry of the KN geometry, the perturbative variable $\delta\phi$ can be decomposed as
 \begin{eqnarray}
 \delta \phi(\tilde{t},r,\theta,\tilde{\varphi}) =\frac{1}{r} \sum_{m} u_m(\tilde{t},r,\theta) e^{i m \tilde{\varphi}}.
\end{eqnarray}
Substituting the above ansatz into Eq.~(\ref{per-eq3}), we can obtain the scalar perturbation equation in the ingoing coordinates
\begin{eqnarray}
&&A^{\tilde{t}\tilde{t}}\partial_{\tilde{t}}^2u_m+A^{\tilde{t}r}\partial_{\tilde{t}}\partial_r u_m+A^{r r}\partial^2_r u_m\nonumber\\
&&+A^{\theta\theta}\partial_\theta^2u_m+B^{\tilde{t}}\partial_{\tilde{t}}u_m
+B^r\partial_r u_m+B^\theta\partial_\theta u_m+C u_m=0\label{phi-eq4}
\end{eqnarray}
with coefficients
\begin{eqnarray}
&&A^{\tilde{t}\tilde{t}}=\rho^2+2Mr-Q^2,~~A^{\tilde{t}r}=2Q^2-4Mr,~~A^{r r}=-\Delta,~~A^{\theta\theta}=-1,\nonumber\\
&&B^{\tilde{t}}=2M-\frac{2Q^2}{r},~~B^r=\frac{2}{r}(a^2+Q^2-Mr)-2i m a,~~B^\theta=-\cot\theta,  \nonumber\\
&&C=\frac{m^2}{\sin^2\theta}-\frac{2(a^2+Q^2-Mr)}{r^2}+\frac{2i m a}{r}+\mu^2_{\rm eff}\rho^2.\label{coeffs}
\end{eqnarray}

As the second step, we need to replace the ingoing Kerr-Schild coordinates $\tilde{t}$ and $r$ by the new time coordinate $T$ and by the compactified radial coordinate $R$. It is worth to point out that R\'{a}cz and T\'{o}th~\cite{Racz:2011qu} (hereafter RT) first introduced the compactified radial coordinate $R$ and the suitable time coordinate $T$  to discuss the late-time behavior of a scalar field on fixed Kerr background. Here we introduce the compactified radial coordinate $R$ and time coordinate $T$ for scalar field perturbation on the KN black hole in the EMS gravity with
\begin{eqnarray}\label{RT1}
\tilde{t}=T+h(R), \quad r=R/\Omega(R),
\end{eqnarray}
where the height function $h(R)$ and conformal
factor $\Omega(R)$ are given by
\begin{eqnarray}\label{RT2}
h(R)=\frac{1+R^2}{1-R^2}-4M \ln(1-R^2), \quad \Omega(R)=\frac{1-R^2}{2}.
\end{eqnarray}
Equations \eqref{RT1} and \eqref{RT2} are the conformal transformation proposed by Moncrief for Kerr black hole except the logarithmic term in the height function~\cite{Racz:2011qu}. According to Ref.~\cite{Zenginoglu:2007jw}, this logarithmic term is required to be present in the scri-fixing gauge of the Kerr black hole. In our case, we also obtain the same forms of the compactified radial coordinate $R$ and time coordinate $T$ for scalar field perturbation on the KN black hole in the EMS gravity.

% Now we need to introduce new suitable compactified radial coordinate $R$ and time coordinate $T$ for scalar field perturbation on the KN black hole in the EMS gravity
% with
% \begin{eqnarray}
% \tilde{t}=T+h(R), \quad r=R/\Omega(R).
% \end{eqnarray}
% Here the height function $h(R)$ and conformal
% factor $\Omega(R)$ is given by
% \begin{eqnarray}
% h(R)=R/\Omega-R-4M \ln\Omega, \quad \Omega(R)=1-R/S,
% \end{eqnarray}
% where $S$ is a free parameter determining both the domain and the foliation.
In addition, we can further define the boost function $H(R)$ with $H(R)=\frac{dh}{dr}(R)$, then we have the relations
 \begin{eqnarray}
 \partial_{\tilde{t}}=\partial_{T},\quad
 \partial_r=-H(R)\partial_{T}+\frac{dR}{dr}\partial_{R}. \label{RT}
 \end{eqnarray}
With Eq.~(\ref{RT}), Eq.~(\ref{phi-eq4}) could be written as
\begin{eqnarray}
&&\partial^2_{T} u_m+\tilde{A}^{T R}\partial_{T} \partial_{R} u_m+\tilde{A}^{RR}\partial^2_R u_m\nonumber\\
&&+\tilde{A}^{\theta\theta}\partial_\theta^2u_m+\tilde{B}^{T}\partial_{T}u_m
+\tilde{B}^R\partial_R u_m+\tilde{B}^\theta\partial_\theta u_m+\tilde{C} u_m=0, \label{phi-eq5}
\end{eqnarray}
where coefficients can be expressed with
\begin{eqnarray}
\{\tilde{A}^{T R},\tilde{A}^{RR},
\tilde{A}^{\theta\theta},\tilde{B}^{T},
\tilde{B}^R,\tilde{B}^\theta,\tilde{C}\}
=\frac{1}{A^{TT}}\{A^{T R},A^{RR},
A^{\theta\theta},B^{T},
B^R,B^\theta,C\}
\end{eqnarray}
and
\begin{eqnarray}
&&A^{TT}=A^{\tilde{t}\tilde{t}}-H A^{\tilde{t} r}+H^2A^{rr},\nonumber\\
&&B^{T}=B^{\tilde{t}}-H B^{r}-\frac{(1-R^2)^2}{2(1+R^2)}H'A^{rr},\nonumber\\
&&A^{RR}=\left(\frac{(1-R^2)^2}{2(1+R^2)}\right)^2A^{rr},\nonumber\\
&&A^{TR}=\frac{(1-R^2)^2}{2(1+R^2)} A^{\tilde{t}r}-\frac{(1-R^2)^2}{1+R^2}HA^{rr},\nonumber\\
&&B^{R}=\frac{(1-R^2)^2}{2(1+R^2)}\left[B^{r}+\left(\frac{(1-R^2)^2}{2(1+R^2)}\right)'A^{rr}\right],
\end{eqnarray}
where the prime denotes the derivative $\frac{d}{dR}$.

Finally, introducing the following auxiliary fields
\begin{eqnarray}
 &&\Psi_m = \partial_R u_m, \\
 &&\Pi_m = \partial_T u_m,\label{phi-eq6}
\end{eqnarray}
one finds the following coupled equations
\begin{eqnarray}
&&\partial_T u_m = \Pi_m,\\
&&\partial_{T} \Psi_m=\partial_T\partial_R u_m=\partial_R \Pi_m,\label{phi-eq7}\\
&&\partial_{T} \Pi_m=-(\tilde{B}^{T} \Pi_m+
\tilde{A}^{TR}\partial_{R} \Pi_m +\tilde{A}^{RR}\partial_{R}\Psi_m
+\tilde{A}^{\theta\theta}\partial_{\theta}^2u_m\nonumber\\
&&\qquad\qquad+\tilde{B}^{R}\Psi_m+
\tilde{B}^{\theta}\partial_{\theta} u_m+\tilde{C} u_m), \label{phi-eq8}
\end{eqnarray}
which are first-order in space and time. After dividing the fields $\{u_m,\Psi_m,\Pi_m\}$ into real and imaginary parts,
the $R$ and $\theta$-differential equations are solved by using the finite difference method,
whereas the time ($T$) evolution is obtained by applying
the fourth-order Runge-Kutta integrator.
Taking into account the computational efficiency, we
compute the equation in a domain $(R_+,1)\times (0,\pi)$ with
grids of $201\times 68$ points in most cases.
Using the RT coordinates leads to the fact that the ingoing (outgoing) boundary conditions at the outer horizon (infinity)
are satisfied automatically.

Now we consider the angular boundary condition. Notice that the coefficients of Eq.~\eqref{phi-eq8} become singular at the pole where $\theta=0$ and $\pi$ in the $\theta$ direction. Following~\cite{Gao:2018acg,Pazos-Avalos:2004uyd}, we can use a staggered grid and add ghost points to implement these conditions. Considering the dependence on the particular azimuthal mode $m$, the angular boundary condition reads as
for $m=0,\pm 2,\ldots$: $u_m(T,R,\theta)=u_m(T,R,-\theta)$ and $u_m(T,R,\pi+\theta)=u_m(T,R,\pi-\theta)$;
for $m=\pm 1,\pm 3,\ldots$: $u_m(T,R,\theta)=-u_m(T,R,-\theta)$ and $u_m(T,R,\pi+\theta)=-u_m(T,R,\pi-\theta)$.

\section{Numerical results}\label{4s}

We choose spherically harmonic Gaussian bells
centered at $R_c$ outside the  horizon as an initial data for $u_{lm}$
\begin{eqnarray}
u_{lm}(T=0,R,\theta)\sim Y_{lm}(\theta)e^{-\frac{(R-R_c)^2}{2\sigma^2}},
\end{eqnarray}
where $Y_{lm}(\theta)$ represents the $\theta$-dependent spherical harmonics and $\sigma$
is the width of the Guassian distribution.
%Because of the relations presented in Eq.\eqref{phi-eq6}, the initial form of $\Pi$ is
%\begin{eqnarray}
%\Pi_{lm}(T=0,R,\theta)=0.
%\end{eqnarray}

\begin{figure}[t!]
\centering
\includegraphics[width=0.5\textwidth]{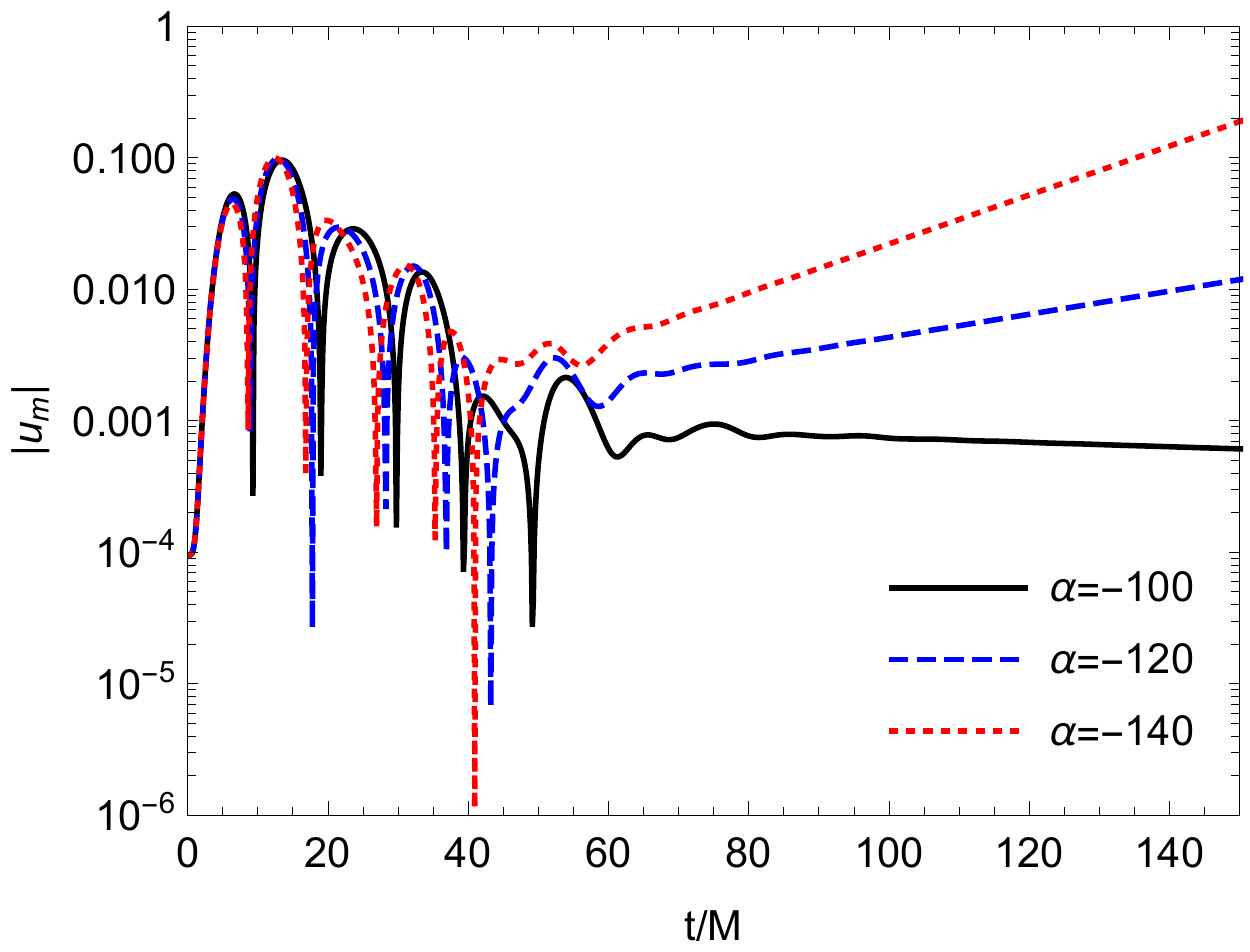}
\caption{ The time-domain profiles of the scalar perturbation for $Q=0.4$, $M=1$ and $a=0.9$ with different $\alpha$. Initial multipole is fixed as
$l=m=0$. }\label{Fig04}
\end{figure}

It is worth to point out that although initially there is only one mode with a specified $l$ number, other $l$ modes with the same index $m$ will be activated during the process of evolution, and the $l=m$ mode will have the dominant contribution at late times if we limit ourselves to stable modes. Similar phenomenon has also occurred for Kerr black hole \cite{Doneva:2020nbb,Gao:2018acg}.
Therefore, in the following we will only consider axisymmetric perturbations with $l=m=0$ for simplicity. In Fig.~\ref{Fig04}, we plot the time-domain profiles of the dominant $l=m=0$ mode of scalar perturbation around KN black holes in the EMS gravity with different values of coupling constant $\alpha$.

\begin{figure}[htbp]
\centering
\subfigure[$Q=0.1$]{
\includegraphics[width=0.45\textwidth]{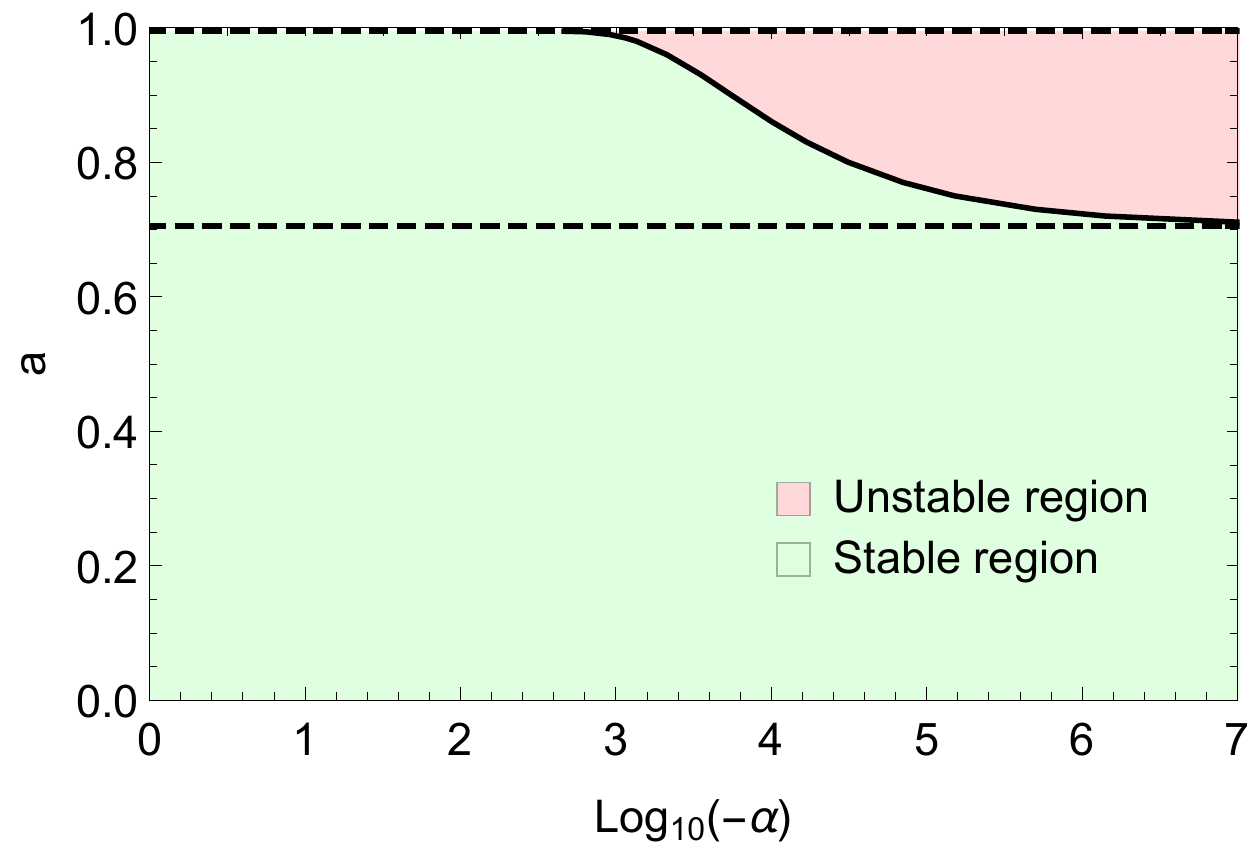}\label{Fig5a}}
\quad
\subfigure[$Q=0.4$]{
\includegraphics[width=0.45\textwidth]{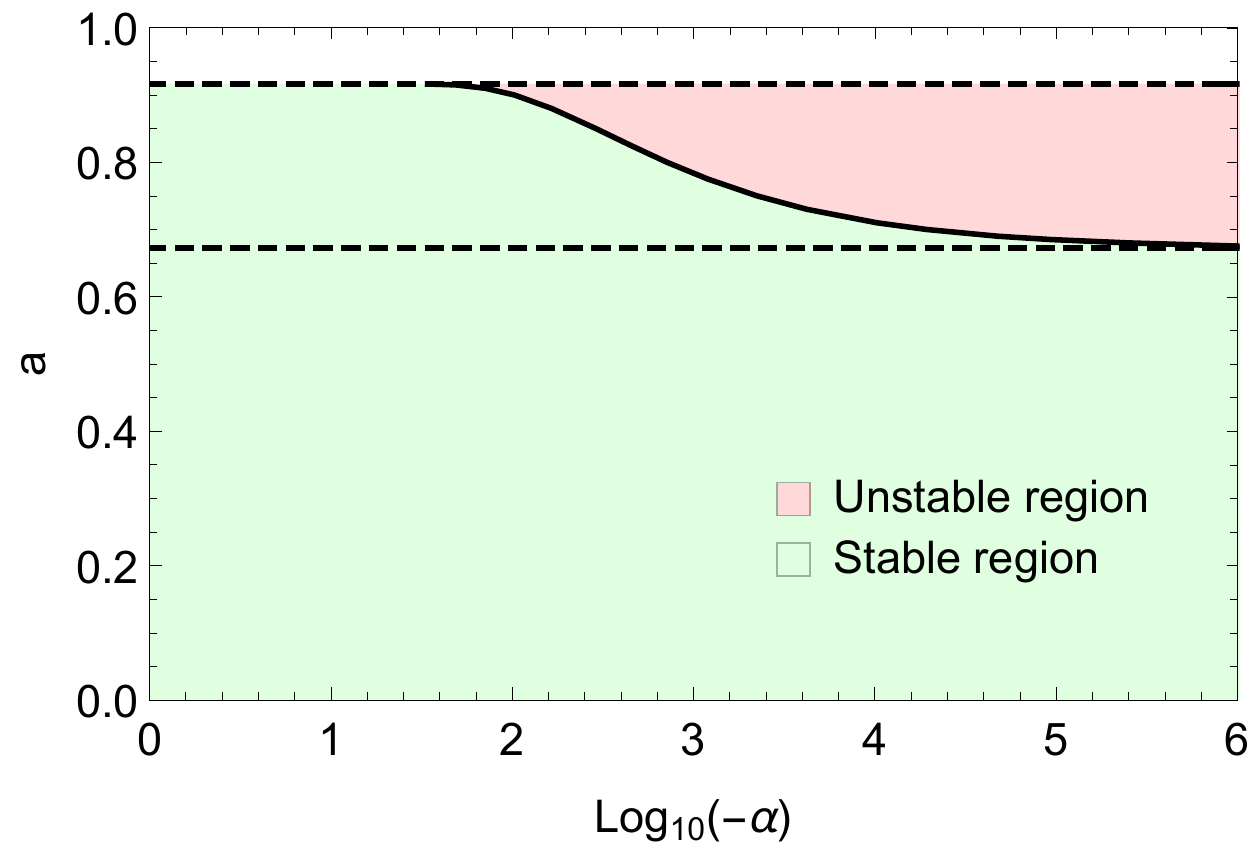}\label{Fig5b}}
\quad
\subfigure[$Q=0.7$]{
\includegraphics[width=0.45\textwidth]{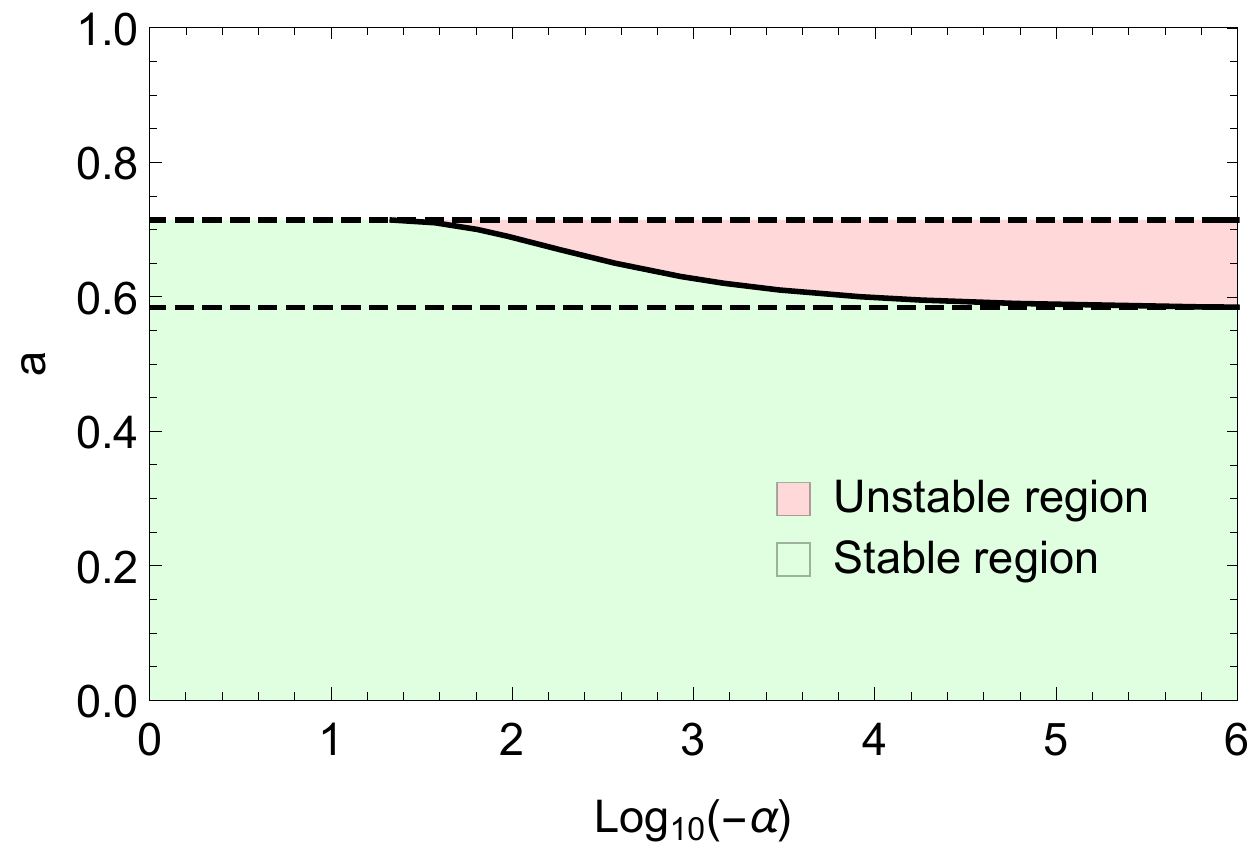}\label{Fig5c}}
\quad
\subfigure[$\frac{a}{r_+}\sim Log_{10}(-\alpha)$]{
\includegraphics[width=0.45\textwidth]{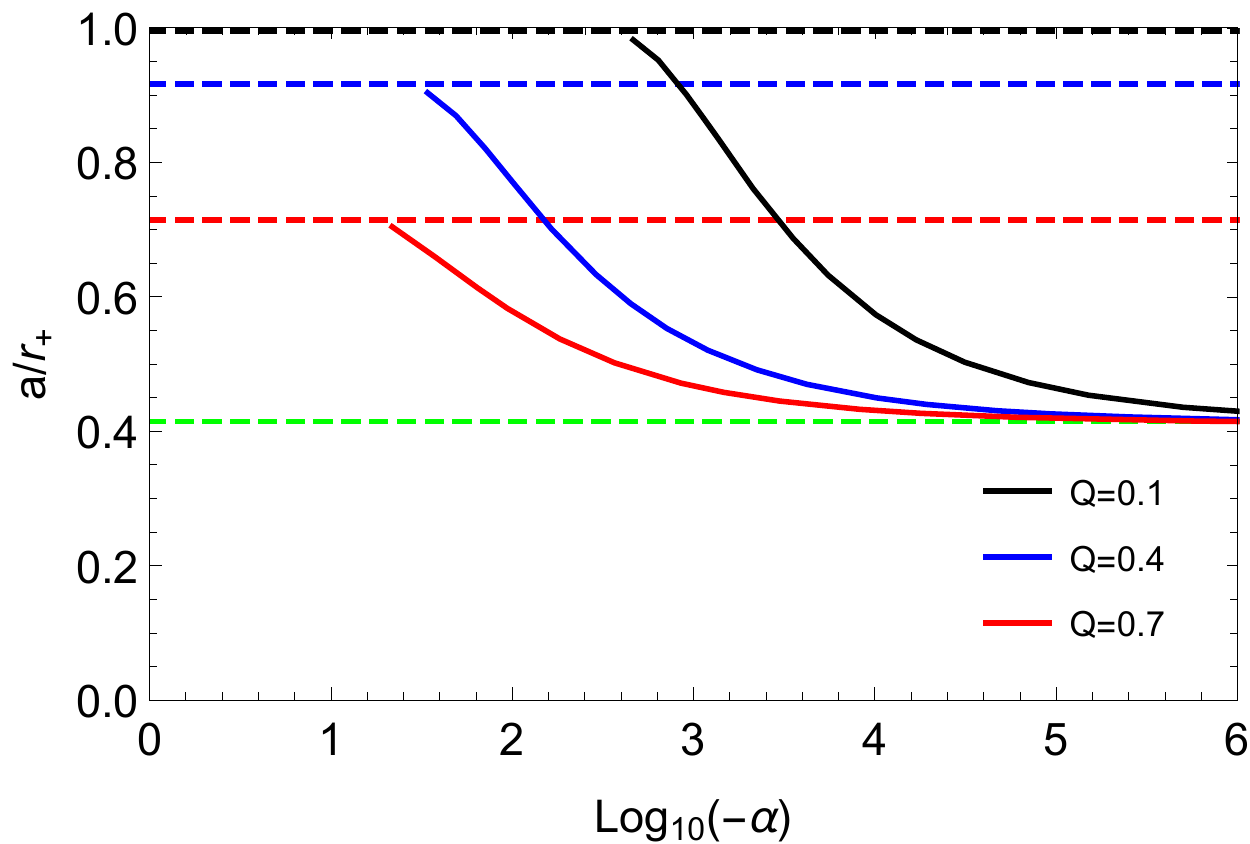}\label{Fig5d}}
\caption{The parameter space for the dominant $m=0$ mode with $Q=0.1, 0.4, 0.7$, and $M=1$. The black and colored solid lines denote the boundaries between stable and unstable region. The black and colored dashed lines represent the upper and lower bounds of $a$ or $a/r_+$.}\label{Fig05}
\end{figure}

From the result above, it is found that the occurrence of
instability depends on the value of spin parameter $a$ and
coupling constant $\alpha$. That is to say, this instability occurs only in a certain region of the parameter space. In Figs.~\ref{Fig5a}-\ref{Fig5c}, we
plot $a$ dividing line in the parameter space spanned by $\alpha$ and $a$ for the $m=0$ mode. For the points in the region above
this line (in red), there exist unstable perturbation modes,
and then the KN black hole becomes unstable, while for the points in
the region below the line (in green), there is no growing mode,
and the KN black hole is stable.

From the numerical results, we further recover the existence of upper and lower bounds for rotation parameter $a$ of KN black hole from Fig.~\ref{Fig05}. Here the upper bound occurs because of the physical constraint $M^2-(a^2+Q^2)\ge 0$ for outer horizon of KN black hole $r_+=M+\sqrt{M^2-(a^2+Q^2)}$. For example, the maximum value of rotation parameter $a$ equals to 0.917 when we set the mass $M=1$ and charge $Q=0.4$ of KN black hole. Moreover, the upper bound of $a$ decreases with the growth of charge $Q$ with fixed $M=1$.

In the meantime, the lower bound of rotation parameter $a$ has been analytically derived in Sec.~\ref{5s} ($a/r_+\ge0.4142$). We find that the gap between the upper and lower bounds becomes narrower as increase of charge $Q$ with fixed mass $M$, see Figs.~\ref{Fig5a}-\ref{Fig5c}. By transforming $a$ to $a/r_+$, we replot the curves of parameters $a\sim\alpha$ with different charge $Q$ of KN black hole in Fig.~\ref{Fig5d}. Then this trend becomes more apparent. As increase of $Q$, a natural deduction is that the bounds will merge together at a particular value of $Q$. This critical charge $Q_{crit}$ can be easily calculated by combining $M^2-(a^2+Q^2)= 0$ and $a/r_+=0.4142$, and the result is $Q_{crit}=0.9102$ with $M=1$. The above argument suggests that the unstable region no longer exists when $Q\ge Q_{crit}$.

\begin{figure}[htbp]
\centering
\subfigure[$\alpha=-103$]{
\includegraphics[width=0.43\textwidth]{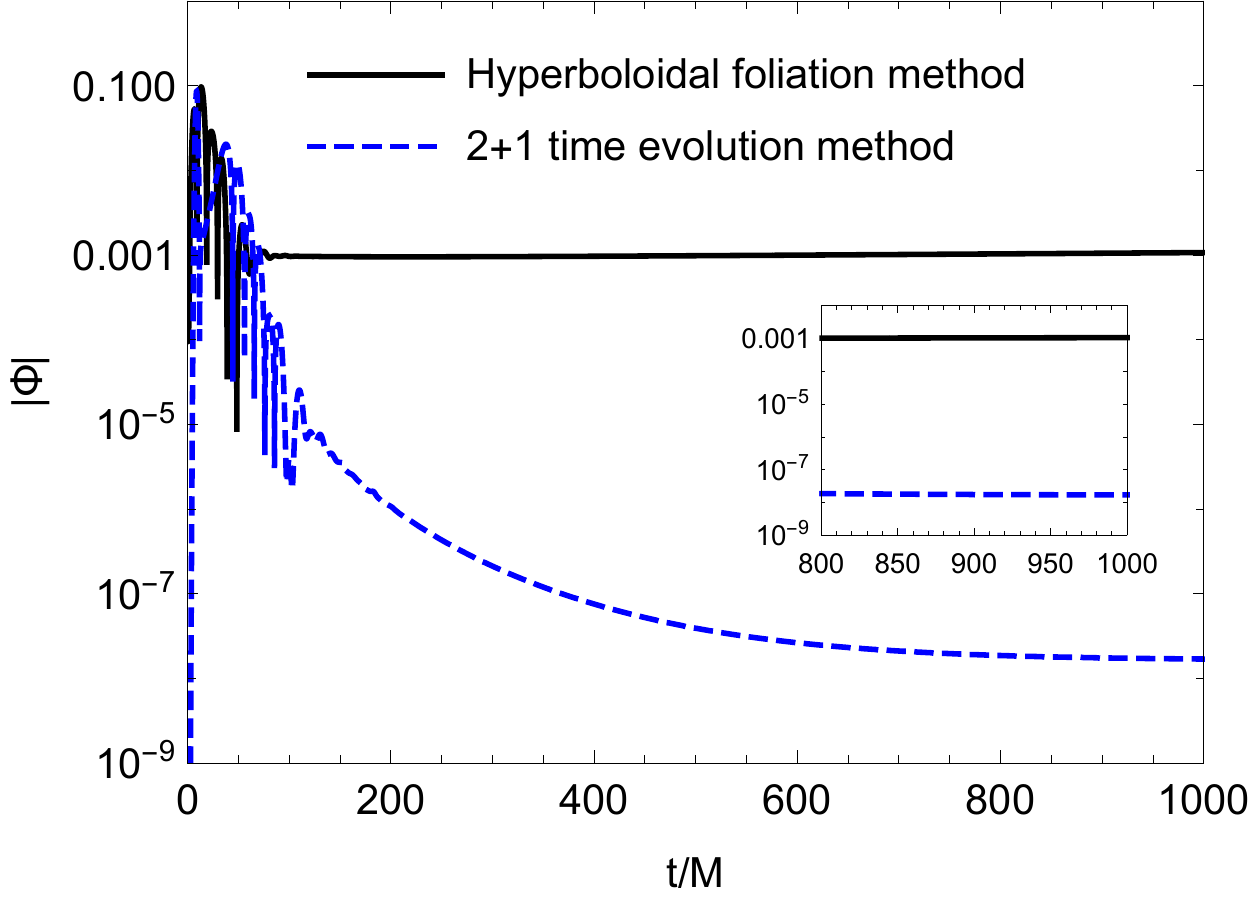}\label{Fig6a}}
\quad
\subfigure[$\alpha=-140$]{
\includegraphics[width=0.43\textwidth]{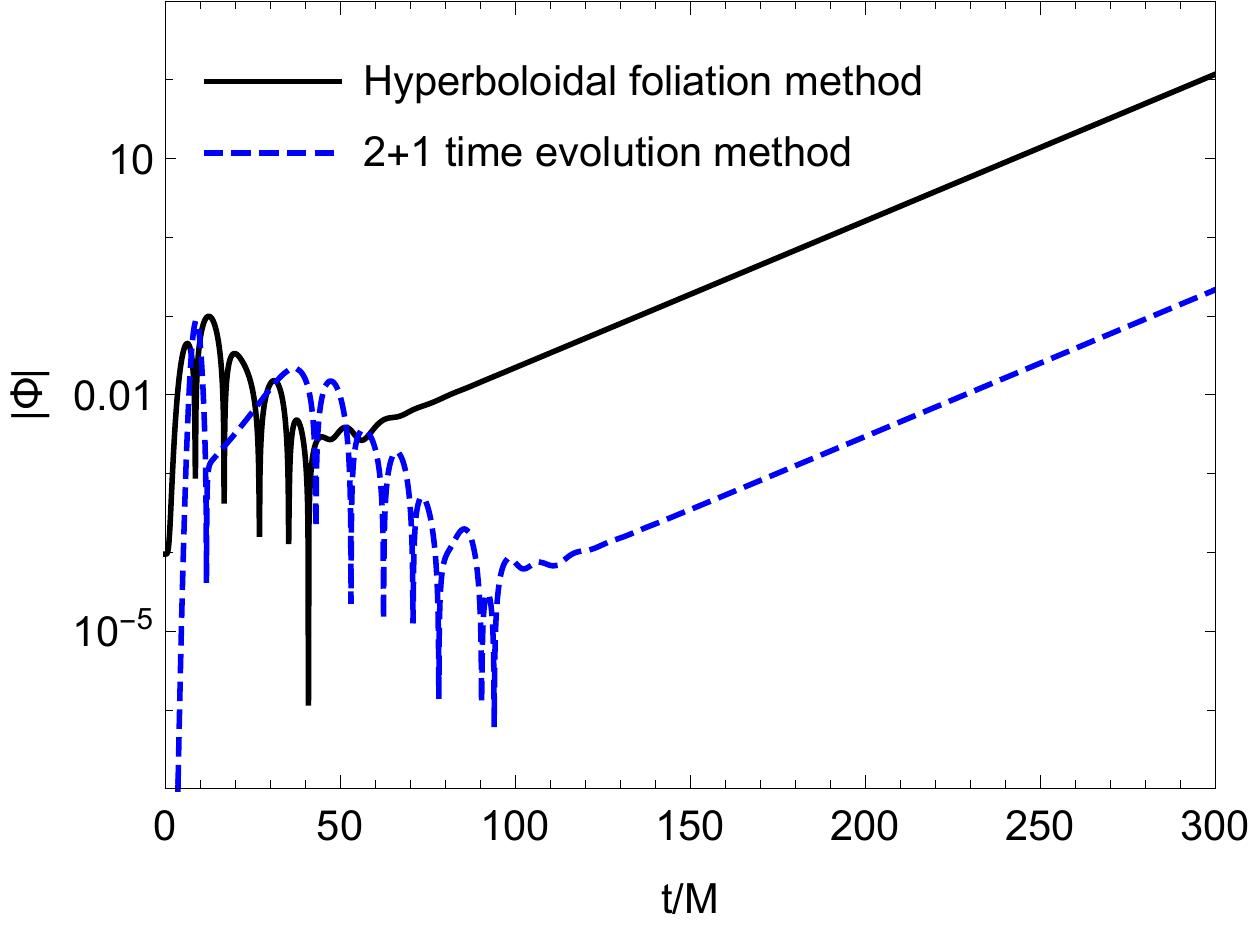}
\label{Fig6b}}
\caption{The time-domain profiles $(m=0)$ of scalar perturbation with $Q=0.4$, $M=1$, $a=0.9$, and different values of $\alpha$ by using two different numerical methods. This figure implies that
the late-time tails of perturbed scalar fields share the same behaviors.}\label{Fig06}
\end{figure}

In the Appendix, we also try to adopt the 2+1 time evolution method \cite{Doneva:2020nbb} to recalculate the late-time tails of perturbed scalar field to perform the tachyonic instability of the KN black holes numerically in time domain. The Fig.~\ref{Fig06} of late-time tails demonstrates that it serves as an independent check of the previous results in last sections.

\section{Summary and discussions}\label{6s}

In this work, we have carried out the tachyonic instability of KN black holes regarding as the onset of spontaneous scalarization in the EMS theory with nonminimal negative scalar coupling to Maxwell term. We firstly used
analytical techniques to obtain an $a$ bound $(\frac{a}{r_+}\geq0.4142)$ in the limit of $\alpha\rightarrow-\infty$, which marks the boundary between
bald KN black holes and hairy (scalarized) spinning black
holes in the EMS gravity.

On the other hand, in order to overcome the nonseparability
problem and the outer boundary problem, we have adopted the hyperboloidal compactification technique to calculate the $(2+1)$-dimensional time evolution of linearized scalar field perturbation on the KN black hole background numerically.  We have got object pictures on the time evolution of the scalar field perturbations. The numerical results reflected nonlinear
effects are expected to quench that instability and lead to a KN black hole with scalar hair. Moreover, we plotted the threshold curve $\alpha(a)$, which describes a boundary between stable and unstable KN black holes. We found that $\alpha(a)$ is a rapidly
decreasing function of $a$ and the high rotation
enhances spontaneous scalarization. This implies that the Maxwell coupling term in the KN black hole plays a similar
role as the GB coupling term in the Kerr black hole background.
It is interesting to note that the analytically derived critical
black hole rotation parameter Eq.~\eqref{critac} agrees remarkably well with the corresponding numerically computed critical rotation parameter (Fig.~\ref{Fig05}).

In this paper, we have concentrated ourselves on the
scalar field perturbation with nonminimal negative scalar coupling to Maxwell term. It is worth to point out that the effective mass term  $\mu^2_{\rm eff}$ in Eq.~\eqref{effmass} is odd  under a combined transformation of $\alpha\to -\alpha$ and $\theta\to\pi-\theta$.
It means that  positive and negative $\alpha$  will induce  different results for tachyonic instability of KN black hole in the EMS gravity. Hereafter, we will consider the spontaneous scalarization phenomenon for KN black hole under the linearized scalar field perturbations with positive coupling parameter $\alpha$ in future work.

 \vspace{1cm}

This work is supported by National Key R$\&$D Program of
China (Grant No. 2020YFC2201400).
D. C. Zou acknowledges financial support from Outstanding young teacher programme from Yangzhou University, No. 137050368. M. Y. Lai acknowledges financial support from the Initial Research Foundation of Jiangxi Normal  University.

 \vspace{1cm}

\newpage

\appendix
\section*{Appendix}

We will calculate the late-time tails of a perturbed scalar field to perform the tachyonic instability of the KN black holes numerically in time-domain. For the linearized scalar equation \eqref{per-eq3}, we introduce a new coordinate $\varphi^*$ and tortoise coordinate $x$ through the transformations~\cite{Zhang:2021btn}
\begin{eqnarray}
  d\varphi^* &=& d\varphi +\frac{a}{\Delta}dr, \nonumber\\
  dx &=& \frac{r^2+a^2}{\Delta}dr. \label{new-coor}
\end{eqnarray}
Then we have a scalar field perturbed equation
\begin{eqnarray}
  &&\left[ (r^2+a^2)^2-
      \Delta a^2\sin^2\theta\right]\partial^2_t \delta\phi
    -(r^2+a^2)^2\partial_x^2\delta\phi
    -2r\Delta\partial_x\delta\phi
   \nonumber\\
    &&+2a\left(2Mr-Q^2\right)\partial_t\partial_{\varphi^*}\delta\phi-2a(r^2+a^2)\partial_x\partial_{\varphi^*}\delta\phi
   -\frac{\Delta}{\sin\theta}\partial_\theta(\sin\theta\partial_\theta\delta\phi)
    \nonumber  \\
    &&-\frac{\Delta}{\sin^2\theta}\partial^2_{\varphi^*}\delta\phi+\Delta\left(r^2+a^2\cos^2\theta\right)\mu_{\rm eff}^2\delta\phi=0,  \label{mscalar-eq}
\end{eqnarray}
where the coordinate $x\in(-\infty,\infty)$ covers the infinite range which  is accessible to an observer located outside the outer horizon, while one notes the semi-infinite region  $r\in[r_+,\infty)$ when using the radial coordinate $r$.

Taking into account the axial symmetry of \eqref{KN-sol}, the scalar perturbation could be decomposed as
\begin{equation}
\delta\phi(t,x,\theta,\varphi^*)=\sum_{m}\delta\phi(t,x,\theta)e^ {im\varphi^*} \label{s-dec}
\end{equation}
with $m$ an azimuthal number.
Substituting \eqref{s-dec} into \eqref{mscalar-eq}, we have a (2+1)-dimensional equation
\begin{eqnarray}
  &&\left[ (r^2+a^2)^2-\Delta a^2\sin^2\theta\right]\partial^2_t\delta\phi -(r^2+a^2)^2\partial^2_x\delta\phi
  -\Delta\partial^2_\theta\delta\phi\nonumber\\
  &&+ 2ima{(2Mr-Q^2)}\partial_t\delta\phi-2\left[r\Delta+ima(r^2+a^2)\right]\partial_x\delta\phi-\Delta\cot{\theta}\partial_\theta\delta\phi \nonumber\\
  &&+  \Delta\Big[(r^2+a^2\cos^2\theta)\mu_{\rm eff}^2+\frac{m^2}{\sin^2\theta}\Big]\delta\phi= 0. \label{mscalar-eq2}
\end{eqnarray}
We may  rewrite \eqref{mscalar-eq2} as the (2+1)-dimensional Teukolsky equation
\begin{eqnarray}\label{perturbedEq8-2}
  \partial^2_t\delta\phi + A^{xx}\partial^2_x\delta\phi
  +A^{\theta\theta}\partial^2_\theta\delta\phi  +B^{t}\partial_t\delta\phi + B^{x}\partial_x\delta\phi+ B^{\theta}\partial_\theta\delta\phi +C \delta\phi= 0
\end{eqnarray}
whose coefficients take the forms
\begin{eqnarray}\label{coeff8-0}
  A^{tt} &=& \left[ (r^2+a^2)^2-\Delta a^2\sin^2\theta\right] , \nonumber\\
  A^{xx} &=&-\frac{(r^2+a^2)^2}{A^{tt}},
  \nonumber\\
  A^{\theta\theta} &=&-\frac{\Delta}{A^{tt}},  \nonumber\\
  B^{t} &=& \frac{ 2ima{(2Mr-Q^2)}}{A^{tt}},
  \nonumber\\
  B^{x} &=& -\frac{ 2r\Delta+2ima(r^2+a^2)}{A^{tt}},
  \nonumber\\
  B^{\theta} &=&-\frac{ \Delta\cot{\theta}}{A^{tt}},
  \nonumber\\
  C &=& \frac{\Delta}{A^{tt}}\left[(r^2+a^2\cos^2\theta)\mu_{eff}^2+\frac{m^2}{\sin^2\theta}\right].
\end{eqnarray}
 We note that for $Q=0$, Eq.~\eqref{perturbedEq8-2} reduces exactly to Eq. (12) in Ref.~\cite{Zhang:2021btn}.
At this stage, we introduce the three auxiliary fields defined by
\begin{eqnarray}
  \Phi &\equiv& \delta\phi, \nonumber\\
  \Psi &\equiv& \partial_x \Phi, \nonumber\\
  \Pi  &\equiv& \partial_t \Phi. \label{aux-f}
\end{eqnarray}
Then, Eq.~\eqref{perturbedEq8-2} can be rewritten as
\begin{eqnarray}\label{perturbedEq8-3}
  \partial_t\Pi
  = -\left(  A^{xx}\partial_x\Psi
  +A^{\theta\theta}\partial^2_\theta\Phi
  +B^t\Pi + B^x\Psi +B^\theta\partial_\theta\Phi+C\Phi\right).
\end{eqnarray}

Dividing the fields into real and imaginary parts
\begin{eqnarray}
% \nonumber % Remove numbering (before each equation)
  \Phi =\Phi_R + i\Phi_I,\quad
  \Psi= \Psi_R + i\Psi_I,\quad
  \Pi = \Pi_R + i\Pi_I,
\end{eqnarray} \label{ri-sep}
Eq.~\eqref{perturbedEq8-3} is separated into two equations
\begin{eqnarray}
 \partial_t\Pi_R=&-&\Big( A^{xx}\partial_x\Psi_R
  +A^{\theta\theta}\partial^2_\theta\Phi_R \nonumber \\
&-&B^t_I\Pi_I + B^x_R\Psi_R- B^x_I\Psi_I+B^\theta\partial_\theta\Phi_R+C\Phi_R\Big),
  \label{eqsv-1}\\
 \partial_t\Pi_I=&-&\Big(  A^{xx}\partial_x\Psi_I
  +A^{\theta\theta}\partial^2_\theta\Phi_I \nonumber \\
&+&B^t_I\Pi_R + B^x_I\Psi_R+ B^x_R\Psi_I +B^\theta\partial_\theta\Phi_I+C\Phi_I\Big).\label{eqsv-2}
\end{eqnarray}
Introducing  $u= (\Phi_R,\Phi_I,\Psi_R,\Psi_I,\Pi_R,\Pi_I)^T$, these equations can be rewritten  compactly as
\begin{eqnarray}
  \partial_t u &=& (G\partial_x+Y)u, \label{comp-eq}
\end{eqnarray}
where
\begin{eqnarray}\label{perturbedEq3}
  G &=&  \left(
           \begin{array}{cccccc}
             0 & 0 & 0 & 0 & 0 & 0 \\
             0  & 0 & 0 & 0 & 0 & 0 \\
             0 & 0 & 0 & 0 & 1 & 0 \\
             0 & 0 & 0 & 0 & 0 & 1 \\
             0 & 0 & G_{53} & 0 & 0 & 0 \\
             0 & 0 & 0 & G_{64} & 0 & 0 \\
           \end{array}
         \right),
  \label{G-mat}\\
   Y &=&  \left(
           \begin{array}{cccccc}
             0 & 0 & 0 & 0 & 1 & 0 \\
             0  & 0 & 0 & 0 & 0 & 1 \\
             0 & 0 & 0 & 0 & 0 & 0 \\
             0 & 0 & 0 & 0 & 0 & 0 \\
             Y_{51} & 0 & Y_{53} & Y_{54} & 0 & Y_{56} \\
             0 & Y_{62}  & Y_{63}  & Y_{64}  & Y_{56}  & 0 \\
           \end{array}
         \right) \label{Y-mat}
\end{eqnarray}
with matrix elements
\begin{eqnarray}
  G_{53} &=& G_{64}=-A^{xx}, \nonumber \\
  Y_{51} &=& Y_{62}= -(A^{\theta\theta}\partial_\theta^2+B^\theta\partial_\theta+C),
  \nonumber\\
  Y_{53} &=& Y_{64}=B^x_R, \nonumber\\
  Y_{54} &=& -Y_{63}=-B^x_I,
  \nonumber\\
  Y_{56}&=&Y_{65}=B^t_I. \label{mat-coeff}
\end{eqnarray}
The derivatives in $x$ and $\theta$ directions are approximated by making use of a finite difference method,
while the time evolution is carried out by adopting the fourth-order Runger-Kutta integrator.
We introduce the boundary conditions: ingoing waves at the outer horizon ($x=-\infty$) and outgoing waves at infinity ($x=\infty$).
At the poles of $\theta=0,\pi$, one impose the boundary condition of $\Phi|_{\theta=0,\pi}=0$   for $m\not=0$, whereas $\partial \Phi|_{\theta=0,\pi}=0$   for $m=0$.

\end{document}